\documentclass{article}
\usepackage{PRIMEarxiv}

\usepackage[utf8]{inputenc} 
\usepackage[T1]{fontenc}    
\usepackage{hyperref}       
\usepackage{url}            
\usepackage{booktabs}       
\usepackage{amsfonts}       
\usepackage{nicefrac}       
\usepackage{microtype}      
\usepackage{lipsum}
\usepackage{fancyhdr}       
\usepackage{graphicx}       
\graphicspath{{media/}}     
\usepackage{multirow}
\usepackage{subcaption}
\usepackage{amsmath,algorithm,amsfonts}
\usepackage{algorithmic}
\usepackage{xspace}
\usepackage{graphicx}
\usepackage{balance}

\title{Dual Adversarial Perturbators Generate rich Views for Recommendation
\thanks{\textit{\underline{Citation}}: 
\textbf{Authors. Title. Pages.... DOI:000000/11111.}} 
}

\author{
  Lijun Zhang \\
  Nanjing University of Aeronautics and Astronautics \\
  Nanjing, China \\
  \texttt{lijunzhang@nuaa.edu.cn}
  \And
  Yuan Yao \\
  Nanjing University \\
  Nanjing, China \\
  \texttt{y.yao@nju.edu.cn}
  \And
  Haibo Ye \\
  Nanjing University of Aeronautics and Astronautics \\
  Nanjing, China \\
  \texttt{yhb@nuaa.edu.cn}
}

\newcommand{\name}{AvoGCL\xspace}
\begin{document}

\maketitle

\begin{abstract}
Graph contrastive learning (GCL) has been extensively studied and leveraged as a potent tool in recommender systems. Most existing GCL-based recommenders generate contrastive views by altering the graph structure or introducing perturbations to embedding. While these methods effectively enhance learning from sparse data, they risk performance degradation or even training collapse when the differences between contrastive views become too pronounced. To mitigate this issue, we employ curriculum learning to incrementally increase the disparity between contrastive views, enabling the model to gain from more challenging scenarios. In this paper, we propose a dual-adversarial graph learning approach, \name, which emulates curriculum learning by progressively applying adversarial training to graph structures and embedding perturbations. Specifically, \name construct contrastive views by reducing graph redundancy and generating adversarial perturbations in the embedding space, and achieve better results by gradually increasing the difficulty of contrastive views. Extensive experiments on three real-world datasets demonstrate that \name significantly outperforms the state-of-the-art competitors.
\end{abstract}


\section{Introduction}
To accomplish personalized information filtering for users, recommender systems find extensive applications across diverse domains including shopping, reading, and movie-watching~\cite{ying2018graph,yuan2020parameter}. 
Within recommender systems, graph collaborative filtering (GCF) has gained widespread adoption due to their proficiency in extracting collaborative signals from the user-item interaction graph. In particular, GCF tends to utilize graph neural networks (GNNs) to unearth the high-order neighborhood information of nodes, facilitating enhanced representation learning for both users and items~\cite{wang2019neural,wang2020disentangled,he2020lightgcn,sun2019multi,chen2020revisiting}.


Recently, contrastive learning has led to a series of successful practices of GCF-based recommenders. We name such methods as {\em graph contrastive learning} (GCL) based recommenders.
Typically, these methods first construct multiple contrastive views for each node, and then pull closer the features of the same node under different views. 
Depending on where the contrastive views are constructed, existing GCL-based recommenders can be roughly divided into two classes. 
The first class constructs contrastive views in the {\em structural space} by perturbating the original user-item interaction graph~\cite{wu2021self,jiang2023adaptive,cai2023lightgcl}.
The second class directly adds random perturbations in the {\em embedding space}~\cite{yu2022graph,yu2023xsimgcl}.

\begin{figure}
    \centering
    \includegraphics[width=0.8\linewidth]{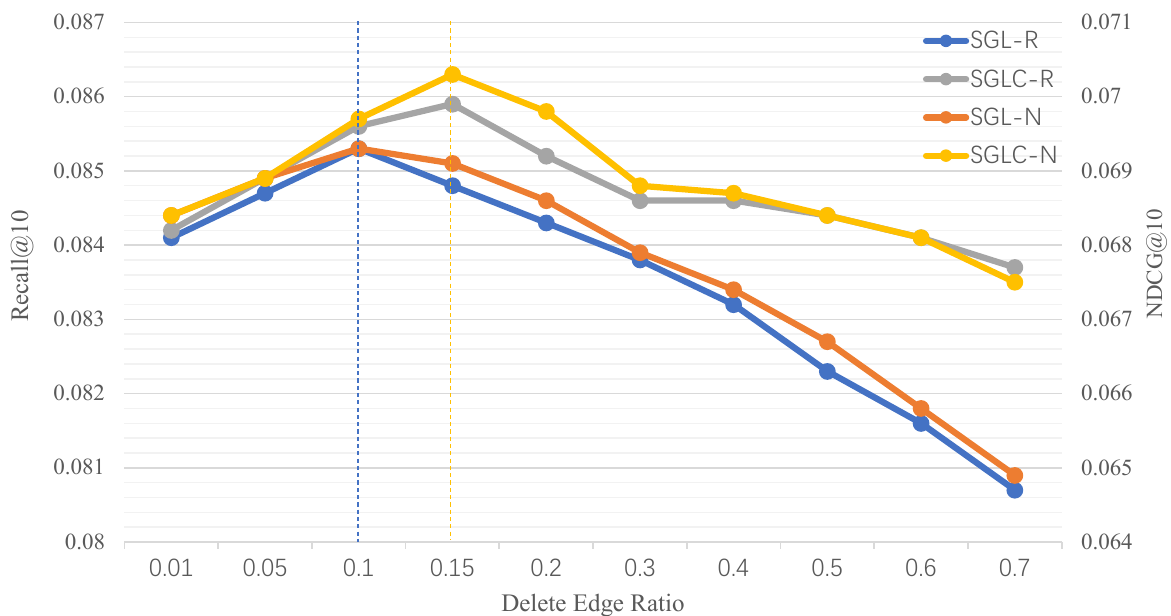}
    \caption{The curves of Recall@10 and NDCG@10 of the model with the edge deletion ratio. SGLC represents the SGL method with added curriculum learning. '-R' represents the Recall@10 score of the model, and '-N' represents the NDCG@10 score of the model.}
    \label{fig:Curriculum_Learning_intro}
\end{figure}

The method of constructing contrastive views through graph structure modification or vector perturbation has achieved notable success. However, an intriguing phenomenon has been observed: increasing the contrastive view difference within a certain range enhances model performance~\cite{tong2021directed}, but excessive differences lead to performance degradation and even training collapse. Take SGL~\cite{wu2021self}, a classic graph contrastive learning method that constructs contrastive views by randomly deleting a proportion of edges, as an example. As shown in Fig.~\ref{fig:Curriculum_Learning_intro}, the blue and orange lines represent SGL's scores on Recall@10 and NDCG@10, respectively. Initially, model performance improves with an increasing edge deletion ratio, peaking at a 0.1 deletion ratio, after which it rapidly declines. Naturally, we add curriculum learning~\cite{wang2021survey} to SGL to gradually increase the ratio of edge deletion and name it SGLC. The gray and yellow lines in Fig.~\ref{fig:Curriculum_Learning_intro} depict SGLC's performance on Recall@10 and NDCG@10, respectively. It is evident that SGLC not only outperforms SGL at the same deletion ratios but also achieves optimal results at a 0.15 deletion ratio. This indicates that generating contrastive views from easy to difficult through curriculum learning can effectively raise the upper bound of contrastive learning.

In this paper, we propose a dual adversarial learning approach to constructs contrastive views with gradually increasing difficulty, named \name, for GCL-based recommendation. 
Instead of random contrastive view construction, \name intentionally constructs more challenging contrastive views in an adversarial way.
Specifically, \name employs two trainable perturbators the {\em structure perturbator} and the {\em embedding perturbator} to generate more targeted graph structures and embedding perturbations. It designs two minimax games: with the former attempting to generate perturbated graph with lower-redundancy compared to the original user-item interaction graph, and the latter aiming to generate higher-variation views that push the embeddings against the contrastive learning loss. This approach ensures that the model encounters contrastive views of progressively increasing difficulty, thereby enhancing its overall performance.

Extensive experiments are conducted on three real-world datasets. The results show that, with the above adversarial construction of contrastive views, \name significantly outperforms the existing competitors. For example, it improves the best existing GCL-based competitors by up to 7.1\% in terms of recommendation accuracy.

Our contributions are highlighted as follows:
\begin{itemize}
\item We propose a new GCL-based recommender that conducts contrastive learning in both the structural space and the embedding space. To the best of our knowledge, this is the first work that combines both in an adversarial learning framework.
\item Drawing on the concept of curriculum learning, we employ adversarial techniques to generate contrastive views of increasing difficulty, thereby pushing the upper bound of contrastive learning. 
\item We propose a lightweight design for an adversarial embedding perturbation generator, which is a plug-and-play solution that ensures effectiveness while maintaining minimal computational overhead.
\item We conduct extensive experiments on three publicly available datasets, and the results show that \name outperforms the current state-of-the-art methods. 
\end{itemize}

The rest of the paper is organized as follows. Section~\ref{sec:pre} introduces the background knowledge. Section~\ref{sec:method} presents the proposed approach, and Section~\ref{sec:exp} shows the experimental results. Section~\ref{sec:rel} covers the related work, and Section~\ref{sec:con} concludes the paper.
\section{Preliminaries}\label{sec:pre}
In this section, we introduce some background knowledge about graph collaborative filtering and graph contrastive learning.

\subsection{Graph Collaborative Filtering}
In a typical GCF task, we have a user set $\mathcal{U}$ and an item set $\mathcal{I}$, along with the observed user-item interactions denoted as graph $G$. We consider the implicit interactions where graph $G$ can be represented by a binary user-items interaction matrix $\mathbf{R}$. 
Each user $u \in \mathcal{U}$ and item $i \in \mathcal{I}$ is assigned a $d$-dimensional embedding, represented as $\mathbf{e}_u$ and $\mathbf{e}_i$, respectively. Given the interaction graph $G$, the purpose of the GCF model is to learn the appropriate embeddings for users and items to achieve accurate recommendations.


Different methods may differ in terms of aggregating the neighborhood information and building the optimization task. Take LightGCN~\cite{he2020lightgcn} as an example. The user and item embeddings in each layer are calculated as follows:
\begin{eqnarray}\label{eq2}
\mathbf{e}_{u}^{l+1}= \sum_{i\in \mathcal{N}_u} \frac{1}{\sqrt{|\mathcal{N}_u||\mathcal{N}_i|}} \mathbf{e}_{i}^{l},  \nonumber\\
\mathbf{e}_{i}^{l+1}= \sum_{u\in \mathcal{N}_i} \frac{1}{\sqrt{|\mathcal{N}_i||\mathcal{N}_u|}} \mathbf{e}_{u}^{l},
\end{eqnarray}
where $\mathcal{N}_u$ and $\mathcal{N}_i$ represent the one-hop neighbors of user $u$ and item $i$, respectively, and we use superscript to indicate the GNN layer. 
After $L$ layers, the embeddings of each layer are merged,
\begin{equation}\label{eq3}
\mathbf{e}_u=\frac{1}{L} \sum_{l=0}^L \mathbf{e}_{u}^{l},\quad \quad \mathbf{e}_i=\frac{1}{L} \sum_{l=0}^L \mathbf{e}_{i}^{l}.
\end{equation}
Then, the preference score of user $u$ towards item $i$ is calculated as the inner product between user embedding and item embedding, i.e.,
$\hat{y}_{u,i}=\mathbf{e}_u^T \mathbf{e}_i$.
Finally, the BPR loss is employed for learning the embeddings of users and items:
\begin{equation}\label{bpr}
\mathcal{L}_{BPR}=\sum_{(u, i, j) \in \mathcal{O}}-\log \sigma\left(\hat{y}_{u,i}-\hat{y}_{u,j}\right),
\end{equation}
where $\mathcal{O}=\left\{(u, i, j) \mid u \in \mathcal{U}, i,j \in \mathcal{I}, {\mathbf{R}}_{u,i} \neq 0, {\mathbf{R}}_{u,j}=0\right\}$, and $\sigma$ denotes the sigmoid function. Essentially, the BPR loss incentivizes predicted preferences of observed interactions to surpass those of unobserved ones. 


\subsection{Graph Contrastive Learning}
Due to its outstanding performance in addressing data sparsity issues, contrastive learning has often been used as an indispensable component of modern recommender systems. Typically, the InfoNCE loss is adopted:
\begin{equation}\label{InfoNCE_user}
\mathcal{L}_{cl}^{user}=\sum_{u \in \mathcal{U}} -\ln \frac{exp(cos(\mathbf{e}_u^{'}, \mathbf{e}_u^{''})/\tau)}{\sum_{k \in \mathcal{U}}exp(cos(\mathbf{e}_u^{'}, \mathbf{e}_k^{''})/\tau)},
\end{equation}
where $\mathbf{e}_u^{'}$ and $\mathbf{e}_u^{''}$ represent the user embeddings from two contrastive views, and $\tau$ is a hyper-parameter, also known as the temperature in the InfoNCE loss. 
The overall loss function for graph contrastive learning (GCL) based recommenders is the combination of graph collaborative filtering and contrastive learning, e.g.,
\begin{equation}\label{loss}
\mathcal{L}_{BPR} + \lambda_1 \cdot (\mathcal{L}_{cl}^{user}+\mathcal{L}_{cl}^{item}) + \lambda_2 \cdot \|\Theta\|_{F}^{2},
\end{equation}
where $\Theta$ denotes the learnable model parameters, $\lambda_1$ controls the contribution of contrastive learning to the model, and $\lambda_2$ controls the $L_2$ regularization strength.

\begin{figure*}
    \centering
    \includegraphics[width=0.9\textwidth]{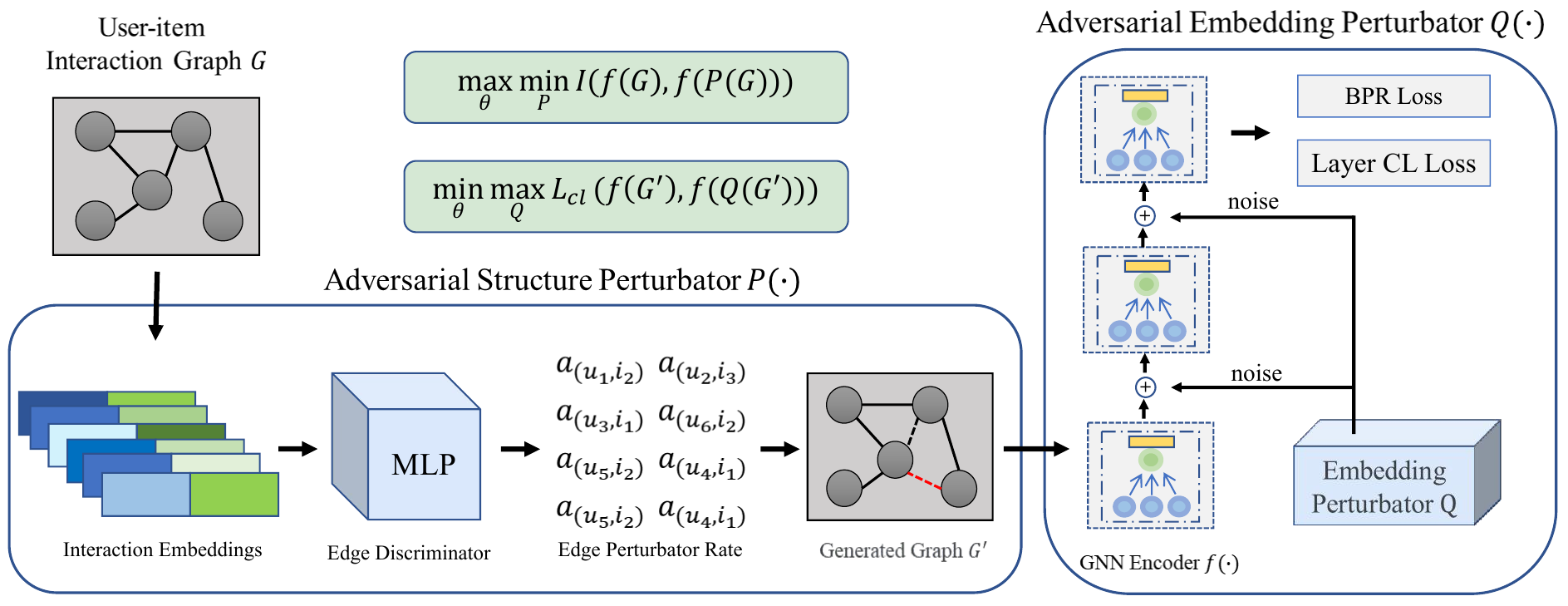}
    \caption{The overview of \name.}
    \label{fig:overview}
\end{figure*}

\section{The Proposed Approach}\label{sec:method}

In this section, we present the proposed approach, \name, whose architecture is illustrated in Fig.~\ref{fig:overview}. \name comprises two principal modules: 1) the {\em adversarial structure perturbator}  and 2) the {\em adversarial embedding perturbator}. 
For the former, it trains a structure perturbator that generates low-redundancy perturbed graphs. 
For the latter, it designs a trainable embedding perturbator that  generates high-variation parameter perturbations. Through adversarial training of these two perturbators, progressively more challenging contrastive views are generated.

\subsection{Adversarial Structure Perturbator}
As mentioned in the introduction, existing  GCL-based recommenders usually adopt random policies to generate perturbated graphs, leaving much room for further improvement.
In this work, we propose to generate lower-redundancy perturbations so as to better unleash the power of  contrastive learning. Specifically, we construct a minimax game: on one hand, we abstract the graph structure perturbation function $P$, which aims to generate perturbated graph $G^{\prime}$ with smaller similarity to the original user-item interaction graph $G$; 
on the other hand, GCL itself tends to bring the two views closer through contrastive learning. 
The adversarial framework is shown below,
\begin{equation}\label{maxmin struct}
\mathop{max}\limits_{\Theta} \mathop{min}\limits_{P}I(f(G),f(P(G))),
\end{equation}
where $I$ represents the similarity function between the two views, $f$ represents the process of GNN aggregation, and $\Theta$ contains the learnable parameters/embeddings under contrastive learning. In the above equation, $G^{\prime} = P(G)$ is the perturbated graph to be learned. Since the maximization problem in Eq.~\eqref{maxmin struct} is the standard contrastive learning, the remaining problem is how to obtain the structure perturbation function $P$. 

{\em Obtaining structure perturbation function P.} Considering the scale and discrete nature of interaction graphs in recommender systems, directly creating a trainable graph generator is costly and time-consuming. Therefore, we propose to use an {\em edge-wise} discriminator to evaluate the {\em importance} of each user-item interaction, deleting/inserting which could maximally distance the perturbated graph from the original one. Specifically,  as shown in Fig.~\ref{fig:overview}, we first randomly sample $\alpha\%$ positive samples (i.e., observed user-item interactions) and the same number of negative samples (i.e., unobserved interactions). 
Then, we estimate the importance of each sample through a discriminator $D$, whose input includes the user/item embeddings and output is a probability indicating the importance of the corresponding edge.
For ease of presentation, we denote the perturbated graph as $G'$, and the sampled positive samples and negative samples as $\mathcal{S}_{pos}$ and $\mathcal{S}_{neg}$, respectively.


Ideally, as stated in Eq.~\eqref{maxmin struct}, training $P$ relies on the similarity function $I$, and the similarity function between two views should be computed based on the outputted user/item embeddings from respective GNN encoders. 
However, due to the discreteness of the interaction graph, back-propagating the loss to the graph structure is non-trivial. 
To this end, inspired by the reconstruction idea of auto-encoders, we propose to reconstruct the samples in $\mathcal{S}_{pos}$ and $\mathcal{S}_{neg}$, and select those with the smallest reconstruction errors as the edges to be deleted/inserted. The intuition is that, edges with smaller reconstruction errors are more possible to be real positive/negative edges, and thus modifying them would make the perturbated graph deviate further from the original one.
Specifically, we adopt the loss function below to train the structure perturbation function $P$:
\begin{equation}\label{loss_mlp}
\sum_{(s_p \in \mathcal{S}_{pos}, ~s_n\in\mathcal{S}_{neg})}{BCE}(D(s_p),1)+{BCE}(D(s_n),0),
\end{equation}
where we use an MLP structure for $D$, and $BCE(\cdot)$ stands for the binary cross-entropy loss. 


During adversarial learning, we iteratively re-sample the positive/negative samples and update discriminator $D$, so that it can continuously generate graphs with reduced similarity to the original graph, introducing adversarial forces against the contrastive learning loss $\mathcal{L}_{cl}$. Simultaneously, the perturbations introduced in the structural space pose additional challenges for the $\mathcal{L}_{BPR}$, further enriching the learning dynamics of the model. 
It is worth mentioning that in order to constrain the loss of graph information within a certain range, we generate $G^{\prime}$ based on the original graph $G$ in each epoch, as opposed to continually generating it based on the previously perturbed graph $G^{\prime}$.


\subsection{Adversarial Embedding Perturbator}
In addition to graph structure perturbation, recent research has found that directly adding random perturbations to the embedding space can produce promising outcomes~\cite{yu2022graph}. 
However, while being viable, the issue of high randomness associated with such methods remains unresolved. Compared to random perturbations, our method aims to stably generate more difficult views, thereby raising the upper limit of contrastive learning.
Similar to the adversarial structure perturbator, we construct an embedding perturbation function $Q$ to add targeted perturbations to each layer of GNNs, leading to more difficult contrastive views. This is also contrary to the goal of contrastive learning which aims to narrow the distance between views. The minimax game is formulated as follows:
\begin{equation}\label{maxmin embedding}
\mathop{min}\limits_{\Theta} \mathop{max}\limits_{Q} \mathcal{L}_{cl}(f(G^{\prime}),f(Q(G^{\prime}))),
\end{equation}
where $f$ still represents the GNN encoder, and $\Theta$ represents the learnable embeddings.
Note that we build the embedding perturbator upon the perturbated graph $G'$.
In the above equation, the minimization problem aims to minimize the contrastive learning loss, while the maximization problem aims to find a $Q$ function that maximizes the same loss.
The remaining problem is how to obtain the $Q$ function. 

{\em Obtaining embedding perturbation function Q.} 
Intuitively, to obtain the embedding perturbations, one can optimize a perturbation matrix for the user/item embeddings in each GNN layer. However, this would be costly especially when there are many users/items or multiple GNN layers. In contrast, we propose to use a projection matrix that can be applied in each GNN layer. Specifically,  taking the user side as an example, we define the projection matrix $\mathbf{K}_{user}$ for users, based on which the embedding perturbation in each layer can be uniformly computed as 
\begin{eqnarray}\label{eq144}
\mathbf{K}_{user}^{'} = \mathbf{E}_{user}\cdot \mathbf{K}_{user}
\end{eqnarray}
\begin{eqnarray}\label{eq145}
\mathbf{Q}_{\text{user}}^{l} = \mathbf{K}_{\text{user}}' \cdot \left(\mathbf{K}_{\text{user}}'^{T} \cdot \mathbf{E}_{\text{user}}^{l}\right)
\end{eqnarray}

In the equation, $\mathbf{K}_{user}$ denotes a $d\times d$ projection matrix, $\mathbf{E}_{user}^{l}$ represents the user embedding at the $l-th$ layer during the current epoch, and $\mathbf{E}_{user}$ stands for the final user embeddings learned from the previous epoch, which serves as our training target. 
Since $\mathbf{E}_{\text{user}}^{l}$ is generated based on the current graph structure $G^{'}$, it may introduce some noise and exhibit locality. Therefore, we first multiply $\mathbf{E}_{user}$ with $\mathbf{K}_{user}$ to obtain the globally informed mapped matrix $\mathbf{K}_{user}^{'}$, which is then multiplied with the current layer's embedding $\mathbf{E}_{\text{user}}^{l}$ to provide targeted perturbations for each layer's embedding. Our later experiment also validates the usefulness of both.

Finally, we add perturbations to the embeddings of each layer and use the perturbated embeddings as the input for the next layer, i.e.,
\begin{equation}\label{eq155}
\mathbf{E}_{user}^{l+1} = (\mathbf{D}^{-1/2} \mathbf{A} \mathbf{D}^{1/2}) \mathbf{E}_{user}^{l} + \omega \cdot \mathbf{Q}_{user}^{l},
\end{equation}
where $\omega$ represents the magnitude of the added perturbation, $\mathbf{A} = [\mathbf{0}, ~~\mathbf{R};~~ \mathbf{R}^T, ~~ \mathbf{0}]$, and $\mathbf{D}$ is the degree matrix of $\mathbf{A}$. 
Similarly, we can optimize $\mathbf{K}_{item}$ for items and obtain the embedding perturbations, which are omitted for brevity.

\subsection{Adversarial Training}
The overall loss function of \name is as follows:
\begin{equation}\label{adversarial learning}
\begin{gathered}
\mathcal{L} = argmin_{\Theta}(\mathcal{L}_{BPR}+\mathcal{L}_{cl})\\
s.t. \quad P^*,Q^* = argmax_{P,Q}(\mathcal{L}_{cl}).
\end{gathered}
\end{equation}
At the training of each epoch, \name generates a low-redundancy perturbated graph $G^{\prime}$ through $P$, calculates the embeddings of users and items based on $G^{\prime}$, and generates contrastive views using the embedding perturbations produced by $Q$. After that, \name updates its parameters $P, Q, \Theta$. In this way, \name adds more vitality to contrastive learning by providing a dissimilar and increasingly challenging adversarial view. The detailed algorithm is summarized in Alg.~\ref{alg:train_algorithm}. 

\begin{algorithm}[t]
\renewcommand{\algorithmicrequire}{\textbf{Input:}}
\renewcommand{\algorithmicensure}{\textbf{Output:}}
\caption{Training algorithm for \name}
\label{alg:train_algorithm}
\begin{algorithmic}[1]
\REQUIRE The original interaction graph $G$, the number of training epochs $T$, hyper-parameters: $\alpha,\omega,\lambda_1,\lambda_2$
\ENSURE The user/item embeddings $\mathbf{E}$
\STATE Initialize parameters $\mathbf{E}$ with Xavier distribution;
\FOR{$t = 1 \rightarrow T$}
    \STATE Generate a perturbated graph $G^{\prime}$ via random sampling; 
    \STATE Optimize the structure perturbation function $P$ via Eq~\eqref{loss_mlp};
    \STATE Calculate the user/item embeddings based on $G^{\prime}$ via Eq.~\eqref{eq2} - Eq.~\eqref{eq3};
    \STATE Calculate the BPR loss $\mathcal{L}_{BPR}$ according to Eq.~\eqref{bpr};
    \STATE Obtain and inject the embedding perturbation based on graph $G^{\prime}$ via Eq.~\eqref{eq144} - Eq.~\eqref{eq155};
    \STATE Calculate the contrastive learning loss $\mathcal{L}_{cl}$  via Eq.~\eqref{InfoNCE_user};
    \STATE Optimize the structure perturbation function $Q$ via Eq.~\eqref{adversarial learning};
    \STATE Calculate joint learning loss of $\Theta$ with Eq.~\eqref{adversarial learning};
    \STATE Backpropagation and update parameters $\mathbf{E}$;
\ENDFOR
\ENSURE $\mathbf{E}=(\mathbf{E}_{user},\mathbf{E}_{item})$
\end{algorithmic}
\end{algorithm}


\subsection{Complexity Analysis}
\begin{table*}[t]
\centering
\caption{The comparison of time complexity.}
\begin{tabular}{crrrrrr}
\hline
    ~ & LightGCN & SGL-ED & SimGCL & XSimGCL & \name\\
\hline
    Adjacency Matrix & $\mathcal{O}(2|A|)$ & $\mathcal{O}((2+4\rho)|A|)$ &  $\mathcal{O}(2|A|)$ & $\mathcal{O}(2|A|)$ & $\mathcal{O}((2+2\alpha)|A|)$\\
    Graph Encoding & $\mathcal{O}(2|A|Ld)$ & $\mathcal{O}((2+4\rho)|A|Ld)$ & $\mathcal{O}(6|A|Ld)$ & $\mathcal{O}(2|A|Ld)$ & $\mathcal{O}((2|A|+N)Ld)$\\
    Prediction & $\mathcal{O}(2Bd)$ & $\mathcal{O}(2Bd)$ & $\mathcal{O}(2Bd)$ & $\mathcal{O}(2Bd)$ & $\mathcal{O}(2Bd)$ \\
    Contrast & - & $\mathcal{O}(BMd)$ & $\mathcal{O}(BMd)$ & $\mathcal{O}(BMd)$& $\mathcal{O}(BMd)$ \\
\hline
\end{tabular}
\label{tab:time complexity}
\end{table*}

Finally, we analyze the time complexity of the proposed method, as well as several existing competitors including LightGCN~\cite{he2020lightgcn}, SGL-ED~\cite{wu2021self}, SimGCL~\cite{yu2022graph}, and XSimGCL~\cite{yu2023xsimgcl}.
The analysis is based on a single batch, and the results are shown in Table~\ref{tab:time complexity}, where we intentionally keep the ordinal numbers for better clarity. We let $|A|$ denote the edge number in the user-item graph, $d$ denote the embedding dimension, $B$ denote the batch size, $M$ denote the node number in a batch, $N$ denote the number of all nodes, $L$ denote the layer number, $\rho$ denote the edge keep rate in SGL-ED, $\alpha$ denote the rate of edge selected by our adversarial structure perturbator.

Compared with the GCL-based method (i.e., SGL-ED) that constructs contrastive views in the structural space, \name is faster ($\rho$ indicates that the rate of undeleted edges, which is usually much larger than $\alpha$). Moreover, since $\alpha$ << 1, the complexity of \name can be regarded the same as that of LightGCN, a GCF-based method that is famous for its lightweight design. 
Compared with the GCL-based methods (i.e., SimGCL and XSimGCL) that construct contrastive views in the embedding space, \name only increases the time complexity of $\mathcal{O}(NLd)$ to achieve control of all perturbations, and $N$ is usually much smaller than $2|A|$. Therefore, the overall design of AvoGCL is also lightweight and efficient.

\section{Experiments}\label{sec:exp}
In this section, we present the experimental results. The experiments are mainly designed to answer the following questions:
\begin{itemize}
    \item RQ1. How does the proposed approach \name perform compared with the existing graph collaborative filtering methods?
    \item RQ2. How does each component of \name contribute to the overall performance? 
    \item RQ3. How does \name perform on noisy and sparse situations?
    \item RQ4. How sensitive is \name w.r.t. its hyper-parameters?
\end{itemize}

\begin{table}[t]
\centering
\caption{Statistics of the experimental datasets.}
\begin{tabular}{crrrr}
\hline
    Dataset & \#User & \#Item & \#Interaction & Sparsity\\
\hline
    Yelp & 45,478 & 30,709 & 1,777,765 & 99.87\% \\
    Gowalla & 29,859 & 40,989 & 1,027,464 & 99.91\% \\
    Amazon & 58,145 & 58,052 & 2,517,437 & 99.92\% \\
\hline
\end{tabular}
\label{tab:data}
\end{table}

\begin{table*}[t]
\centering
\caption{Performance comparison results. The best results are in bold and the second-best  are underlined. \#Improv. represents the improvement of our method compared to the best results of existing competitors. The proposed \name outperforms all the competitors in all cases.}
\begin{tabular}{ccccccccccc}
\hline
Dataset &    Metric &   NGCF &  LightGCN &  SGL &    NCL &  SimGCL &  XsimGCL & LightGCL &  \name &  \#Improv. \\
\hline
\multirow{4}*{Yelp} &   Recall@10 & 0.0584 &    0.0627 & 0.0845 & 0.0847 &  0.0851 &   \underline{0.0853} &   0.0596 & \textbf{0.0892} & 4.57\%
\\~ & Recall@20 & 0.0960 &    0.0996 & 0.1279 & 0.1308 &  \underline{0.1313} &   0.1302 &   0.0992 & \textbf{0.1337} & 1.83\%
\\~ &   NDCG@10 & 0.0444 &    0.0492 & \underline{0.0681} & 0.0680 &  0.0671 &   0.0678 &   0.0596 & \textbf{0.0728} & 7.06\%
\\~ &   NDCG@20 & 0.0568 &    0.0612 & 0.0822 & \underline{0.0831} &  0.0820 &   0.0823 &   0.0728 & \textbf{0.0871} & 4.81\%\\
\hline
\multirow{4}*{Gowalla} & Recall@10 & 0.0974 &    0.1124 & 0.1323 & 0.1348 &  0.1330 &   \underline{0.1354} &    0.1160 & \textbf{0.1432} & 5.76\%
\\~  & Recall@20 & 0.1420 &    0.1616 & 0.1928 & 0.1931 &  0.1879 &   \underline{0.1902} &   0.1695 & \textbf{0.2021} & 6.26\%
\\~ &   NDCG@10 & 0.0776 &    0.0898 & 0.1058 & 0.1064 &  0.1079 &   \underline{0.1102} &   0.0827 & \textbf{0.1166} & 5.81\%
\\~ &   NDCG@20 & 0.0915 &    0.1050 & 0.1240 & 0.1256 &  0.1249 &   \underline{0.1270} &   0.1234 & \textbf{0.1348} & 6.14\%\\
\hline
\multirow{4}*{Amazon} & Recall@10 & 0.0590 &  0.0633 & 0.0869 & 0.0824 &  0.0901 &   \underline{0.0937} &     0.0566 & \textbf{0.0963} & 3.42\%
\\~ & Recall@20 & 0.0931 &    0.0983 & 0.1316 & 0.1211 &  0.1344 &   \underline{0.1367} &     0.0842 & \textbf{0.1423} & 3.37\%
\\~ &   NDCG@10 & 0.0450 &    0.0498 & 0.0689 & 0.0643 &  0.0713 &   \underline{0.0751} &     0.0524 & \textbf{0.0782} & 3.86\%
\\~ &   NDCG@20 & 0.0562 &    0.0612 & 0.0834 & 0.0812 &  0.0856 &   \underline{0.0889} &     0.0657 & \textbf{0.0930} & 3.71\%\\
\hline
\end{tabular}
\label{tab:Normal method gap}
\end{table*}

\subsection{Experimental Setup}
\subsubsection{Datasets} 
We conduct experiments on three widely-used benchmark datasets: Yelp\footnote{https://www.yelp.com/dataset}, Gowalla-merged (Gowalla)~\cite{cho2011friendship},and Amazon Books (Amazon)~\cite{mcauley2015image}. The statistics of these datasets are shown in Table 1. For each dataset, we randomly divide the observed user-item interactions into the training set, validation set, and test set with a ratio of 8:1:1. Following  NCL~\cite{lin2022improving}, we filter out users/items with less than 15 interactions in Yelp and Amazon; for Gowalla, we screen out users/items with less than 10 interactions. We also remove interactions with ratings below 3 in Yelp and Amazon. 

\subsubsection{Evaluation Metrics} 
We evaluate the top-$N$ recommendation performance by using two widely-used metrics: $Recall@N$ and $NDCG@N$, where the value of N is set to \{10, 20\}. We adopt the full-ranking strategy~\cite{zhao2020revisiting}, which ranks all the candidate items that the user has not interacted with. Considering that most methods have randomness, each reported result in the experiment is the average of five runs.

\subsubsection{Compared Methods} 
We compare the proposed \name with the following competitors:
\begin{itemize}
    \item \textbf{NGCF}~\cite{wang2019neural}: It uses a multi-layer graph convolutional network to propagate information through the user-item interaction graph. 
    \item \textbf{LightGCN}~\cite{he2020lightgcn}: This model simplifies the previous work 
    by using a layer-wise propagation scheme that involves only linear transformations and element-wise additions.
    \item  \textbf{SGL}~\cite{wu2021self}: SGL augments LightGCN with self-supervised contrastive learning. It conducts data augmentation through random walk and node/edge dropout to perturb graph structures.
    \item  \textbf{NCL}~\cite{lin2022improving}: NCL enhances graph collaborative filtering by incorporating potential neighbors into contrastive pairs. It introduces structural and semantic neighbors of a user or an item, developing a structure-contrastive and a prototype-contrastive objective.
    \item \textbf{SimGCL}~\cite{yu2022graph}: This model utilizes contrastive learning as an auxiliary task. It performs contrastive learning in the parameter space by adding random noises to the user/item embeddings.
    \item \textbf{CGI}~\cite{wei2022contrastive}:
    CGI is a method that constructs an optimal graph structure by adaptively deleting edges and nodes through information bottlenecks.
    \item \textbf{XSimGCL}~\cite{yu2023xsimgcl}: XSimGCL is an extension of SimGCL, which  changes from the contrast between different graph structures to the contrast between different layers of the same graph structure, greatly reducing the calculation time.
    \item \textbf{LightGCL}~\cite{cai2023lightgcl}:
    LightGCL is a lightweight graph contrastive learning method, which generates contrastive views  through singular value decomposition.
    \item \textbf{LDA-GCL}~\cite{huang2023adversarial}:
    LDA-GCL proposes a learnable graph structure data augmentation in an adversarial way.
\end{itemize}
Among the above competitors, the first two baselines are typical GCF baselines, and the rest are GCL-based methods with good results or research significance. Compared with traditional methods such as BPRMF~\cite{rendle2009bpr} and NeuMF~\cite{he2017neural}, these methods have demonstrated outstanding performance.

\subsubsection{Hyper-parameter Settings} 
For fair comparison, all models are trained from scratch and are initialized with the Xavier distribution. All the models are optimized by the Adam optimizer with learning rate 0.001 and mini-batch size 4096. The early stopping strategy is adopted. For the proposed \name, based on the validation set, we tune the hyper-parameters $\omega$ in \{0.005, 0.01, 0.02, 0.03, 0.04, 0.05\}, $\alpha$ in \{0.01, 0.02, 0.03, 0.04, 0.05\}, and $\lambda_1$ in \{0.1, 0.2, 0.5, 2, 5\}. Other hyper-parameter settings are consistent with XSimGCL. All the experiments were carried out on a server equipped with 24GB RAM, one 16-core Intel i9-9900K CPU@3.60GHz 
and four NVIDIA GeForce RTX 4090 GPUs. For all methods used for comparison,
we reproduce them using RecBole~\cite{zhao2021recbole}.\footnote{We will make the code publicly available.}

\subsection{Overall Performance Comparison (RQ1)}
We first compare \name with the existing competitors, and the results are shown in Table~\ref{tab:Normal method gap}.
\begin{itemize}
    \item It can be first observed that the proposed \name achieves the best results on all the three datasets under both top-10 and top-20 settings.  For example, on the Yelp dataset, \name increases the NDCG@10 metric of the best competitor by 7.1\%; on the  Gowalla dataset, the average increase of the four metrics is 6.0\%.
    \item Second, among the competitors, the GCL-based methods (the latter five competitors) generally outperform the first two competitors of NGCF and LightGCN. This result validates the effectiveness of contrastive learning in GCF.
    \item Third, among the five GCL-based methods, SimGCL and XsimGCL perform generally better than the other three. Note that these two methods construct contrastive views through embedding perturbations. This result indicates the necessarity of embedding perturbations. 
\end{itemize}
We summarize the primary reasons for the superior performance of \name as follows. First, \name constructs more challenging contrastive views for both structural space and embeddings space. 
Second, \name integrates the two adversarial perturbators. 

\subsection{Ablation Study (RQ2)}
We next conduct a ablation study including the following aspects.

\subsubsection{Performance Gain Analysis}
\begin{figure}
    \centering
    \includegraphics[width=0.8\textwidth]{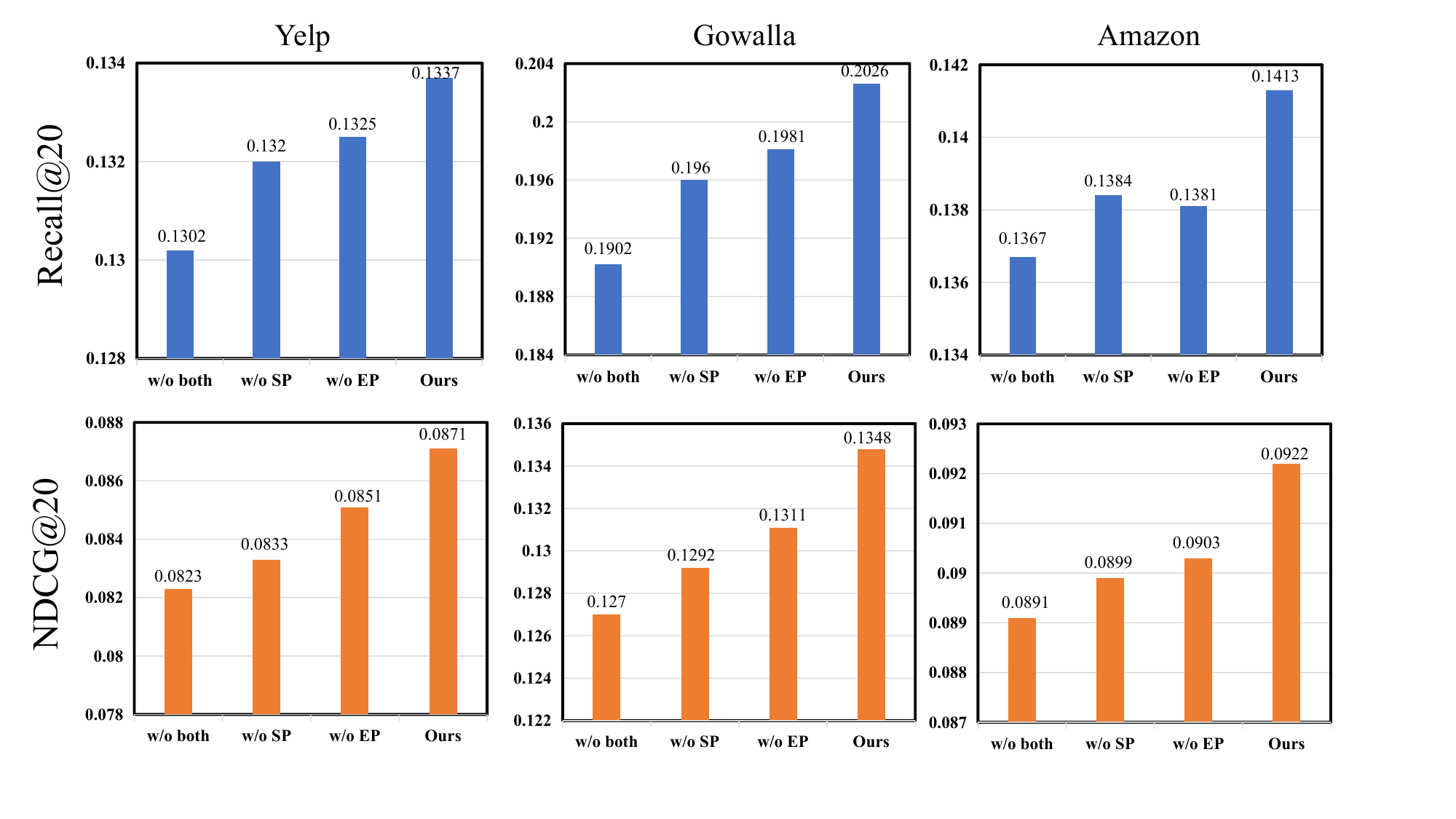}
    \caption{The performance gain analysis of \name. Both the adversarial structure perturbator and the adversarial embedding perturbator can improve the performance of \name.}
    \label{fig:ablation experiments}
\end{figure}

To assess the impact of each component in \name, we next conduct an ablation study by removing each of the two components: the adversarial structure perturbator and the adversarial embedding perturbator. The results are shown in Fig.~\ref{fig:ablation experiments}, where `Ours' represents the proposed \name, `w/o SP' represents \name without the adversarial structure perturbator, `w/o EP' represents \name without the adversarial embedding perturbator, and `w/o both' means deleting both modules. Note that `w/o both' also stands for XSimGCL~\cite{yu2023xsimgcl}, upon which \name is built. It is also the best competitor and thus used as a baseline.

It can be observed that omitting either of the two modules leads to a significant drop in model performance, and discarding the SP module will cause a larger decline. For instance,  on the NDCG@20 metric of the Gowalla dataset, \name without SP, \name without EP, and \name have increased the baseline by 1.73\%, 3.23\%, and 6.14\%, respectively. 
We posit that the relatively modest impact of `w/o SP' can be attributed to the fact that XSimGCL has already used embedding perturbation to generate contrastive views.
Overall, these observations affirm the indispensable roles played by both components in enhancing the model performance.

\begin{table}[t]
\centering
\caption{Performance comparison of XsimGCL variants with different types of disturbances. The adversarial embedding perturbation is better than the random perturbation in XSimGCL. Both components in Eq.~\eqref{eq144} matters.}
\begin{tabular}{c|cc|cc|cc}
\hline
\multirow{2}*{Method} &  \multicolumn{2}{c|}{Yelp} & \multicolumn{2}{c|}{Gowalla} &\multicolumn{2}{c}{Amazon}\\
\cline{2-7}
~ &R@20 &    N@20 &      R@20 &    N@20 &    R@20 &    N@20 \\
\hline
  XSimGCL &    0.1302 &     0.0823 &       0.1902&      0.1270&       0.1367 &     0.0891 \\
   Gaussian &    0.1307 &     0.0823 &         0.1899&     0.1267&       0.1365 &     0.0889 \\
adv\_epoch &    0.1245 &     0.0789 &         0.1815&     0.1212&       0.1345&     0.0876 \\
adv\_layer &    0.1261 &     0.0797 &         0.1769&     0.1179&       0.1337 &     0.0873 \\
EP &     0.1320 &     0.0833 &         0.1960&      0.1292&       0.1384 &     0.0899 \\
\hline
\end{tabular}
\label{tab:embedding perturbation method gap}
\end{table}

\subsubsection{Different Embedding Perturbations}
Next, we compare the performance of various embedding perturbation methods with XSimGCL~\cite{yu2023xsimgcl} as the backbone network, and the results are shown in  Table~\ref{tab:embedding perturbation method gap}. XSimGCL uses random uniform perturbation noise with sign constraints, `Gaussian' represents Gaussian noise, and `EP' represents the noise generated by \name. In Eq.~\eqref{eq144}, we apply the projection matrix on both the embeddings from the previous epoch (i.e., $\mathbf{E}$) and the embeddings of the current layer in the current epoch (i.e., $\mathbf{E}^l$). Here, we also test two variants `$\rm adv\_epoch$' and  `$\rm adv\_layer$', which represent the noise produced by using $\mathbf{E}$ and $\mathbf{E}^l$ only, respectively.

There are several observations. First, by changing the embedding perturbations in XSimGCL with our adversarial one, further accuracy improvement is achieved.
Second, when generating embedding perturbations, using only $\mathbf{E}$ or $\mathbf{E}^l$ does not perform well. The reasons are as follows. If only $\mathbf{E}$ is considered, it leads to the same perturbation for each layer of GNNs, which is undesired. If only $\mathbf{E}^l$ is considered, user/item embeddings can be easily misled by the structure perturbations especially in the early training stage.
Our method integrates both, which not only provides targeted perturbations for each layer of GNNs, but also allows each user/item to obtain a relatively stable embedding vector over iterations. 

\begin{table*}[t]
\centering
\caption{Relation between graph redundancy and model performance. Lower redundancy leads to better performance.}
\begin{tabular}{c|ccc|ccc|ccc}
\hline
\multirow{2}*{Method} &  \multicolumn{3}{c|}{Yelp} & \multicolumn{3}{c|}{Gowalla} &\multicolumn{3}{c}{Amazon}\\
\cline{2-10}
~ &R@20 &N@20 &Sim&R@20 &N@20 &Sim& R@20 &N@20  &Sim \\
\hline
SGL &0.1279 &     0.0822 &     99.77\% &         0.1928 &      0.1240 &     99.68\% &       0.1316 &     0.0834 &     99.76\% \\
SGL\_pos &    0.1308 &     0.0841 &     99.21\% &         0.1968 &     0.1272 &     99.49\% &       0.1347 &     0.0856 &     99.38\% \\
SGL\_both &    0.1315 &     0.0849 &     98.37\% &         0.1979 &     0.1278 &     98.92\% &       0.1352 &     0.0863 &     99.03\% \\
\hline
\end{tabular}
\label{tab:subgraph redundancy}
\end{table*}

\subsubsection{The Impact of Graph Redundancy}
Note that the key idea of our adversarial structure perturbator is to lower the graph redundancy. To examine the impact of graph redundancy on model performance, we apply GNNs on the original graph and the perturbated graph, and compare the cosine similarity of the two output views. We use SGL~\cite{wu2021self} as the backbone network and compare three methods: random edge deletion, discriminator-based edge deletion, and discriminator-based edge addition and deletion (as used in \name). 

The experimental results are shown in Table~\ref{tab:subgraph redundancy}.
We can observer that, as the graph redundancy  decreases, the performance increases. This result also confirms that the proposed structure perturbator can successfully lower the graph redundancy and thus improve the performance.
Also, we find that allowing the model to add edges as well as delete edges results in better performance.


\begin{table*}[t]
\centering
\caption{Performance comparison on datasets with injected noisy interactions. The best results are in bold and the second-best are underlined. \#Improv. represents the improvement of our method compared to the best results of existing competitors. The proposed \name still outperforms all the competitors in all cases.}
\begin{tabular}{ccccccccccc}
\hline
Dataset &    Metric &   NGCF &  LightGCN &  SGL &  NCL &  SimGCL &  XsimGCL & LightGCL &  \name &  Improv. \\
\hline
\multirow{4}*{Yelp-20\%} &   recall@10 & 0.0413 &    0.0577 &  0.0762 & 0.0733 &  0.0797 &    \underline{0.0804} &    0.0443 & \textbf{0.0845} & 5.10\%
\\~ & recall@20 & 0.0693 &    0.0930 &  0.1192 & 0.1141 &   \underline{0.1249} &   0.1248 &    0.0712 & \textbf{0.1264} & 1.28\% 
\\~ &   ndcg@10 & 0.0317 &    0.0448 &  0.0608 & 0.0583 &  0.0628 &   \underline{0.0641} &    0.0499 & \textbf{0.0687} & 7.18\%
\\~ &    ndcg@20 & 0.0408 &    0.0562 &  0.0747 & 0.0713 &  0.0773 &    \underline{0.0785} &    0.0577 & \textbf{0.0822} & 4.71\%\\
\hline
\multirow{4}*{Gowalla-20\%} & recall@10 & 0.0774 &    0.1076 &  0.1266 & 0.1229 &  0.1308 &    \underline{0.1331} &    0.1031 & \textbf{0.1390} & 4.43\%
\\~  & recall@20 & 0.1134 &    0.1547 &  0.1818 & 0.1758 &  0.1862 &    \underline{0.1892} &    0.1522 & \textbf{0.1968} & 4.02\%
\\~ &   ndcg@10 & 0.0625 &    0.0873 &  0.1025 & 0.0989 &  0.1054 &   0.1093 &     \underline{0.1094} & \textbf{0.1137} & 3.93\%
\\~ &   ndcg@20 & 0.0737 &    0.1018 &  0.1196 & 0.1152 &  0.1226 &    \underline{0.1265} &    0.1231 & \textbf{0.1314} & 3.87\%\\
\hline
\multirow{4}*{Amazon-20\%} & recall@10 & 0.0348 &    0.0571 &  0.0772 & 0.0706 &  0.0881 &    \underline{0.0892} &       0.0536 & \textbf{0.0915} & 2.58\%
\\~ & recall@20 & 0.0570 &    0.0887 &  0.1165 & 0.1104 &  0.1281 &    \underline{0.1310} &       0.0779 & \textbf{0.1348} & 2.90\%
\\~ &   ndcg@10 & 0.0271 &    0.0447 &  0.0614 & 0.0593 &  0.0714 &    \underline{0.0716} &       0.0477 & \textbf{0.0738} & 3.07\%
\\~ &   ndcg@20 & 0.0344 &    0.0549 &  0.0740 & 0.0688 &  0.0842 &   \underline{0.0850} &       0.0583 & \textbf{0.0877} & 3.18\%\\
\hline
\end{tabular}
\label{tab:noise method gap}
\end{table*}

\subsection{Robustness Evaluation (RQ3)}
Here, we measure the robustness of \name on noisy and sparse datasets.

\subsubsection{Performance on Noisy Data}
In order to evaluate the impact of noisy interactions on model performance, we randomly injected 20\% edges to the training set and retrain the model. We compare \name with other methods, and the results are shown in Table~\ref{tab:noise method gap}. We observe that \name still achieves the best results in all cases compared to the baselines. For example, on the Yelp dataset, \name increases the NDCG@10 metric of the best competitor by 7.2\%.


\begin{figure}
    \centering
    \begin{subfigure}{0.8\textwidth}
        \centering
        \includegraphics[width=\textwidth]{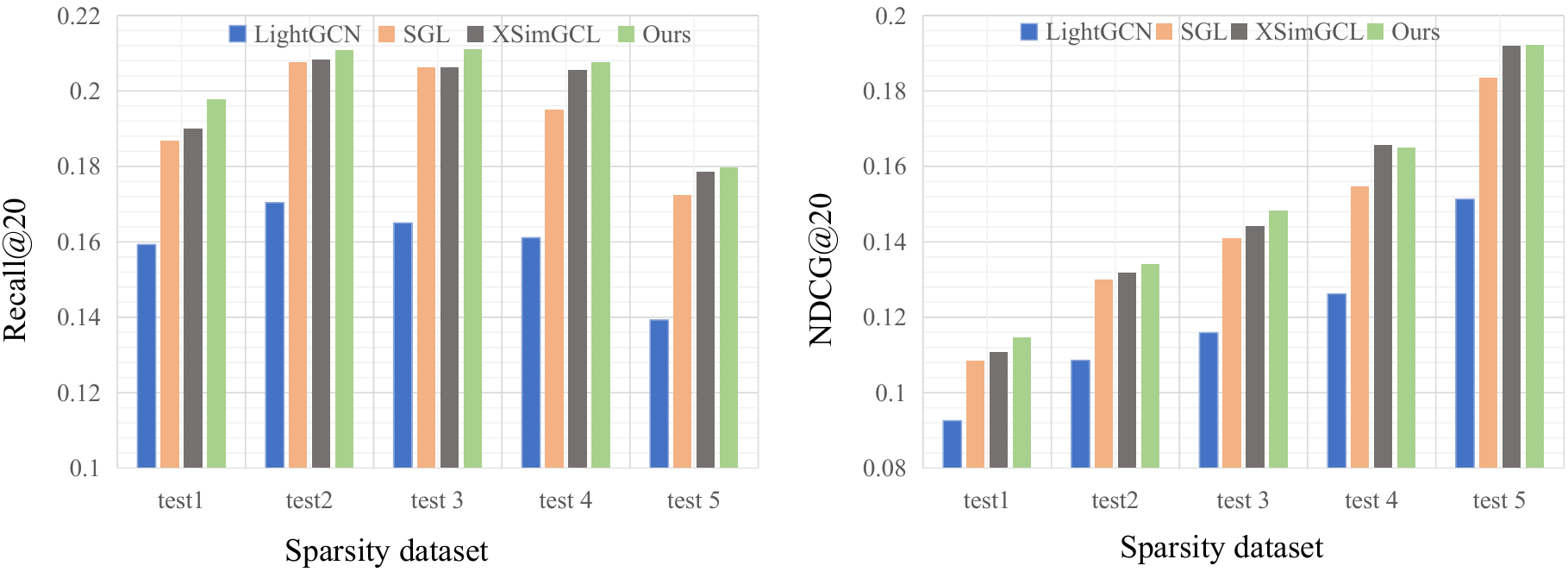}
        \caption{Performance with user interaction numbers}
        \label{fig:sub1}
    \end{subfigure}
    \par
    \begin{subfigure}{0.8\textwidth}
        \centering
        \includegraphics[width=\textwidth]{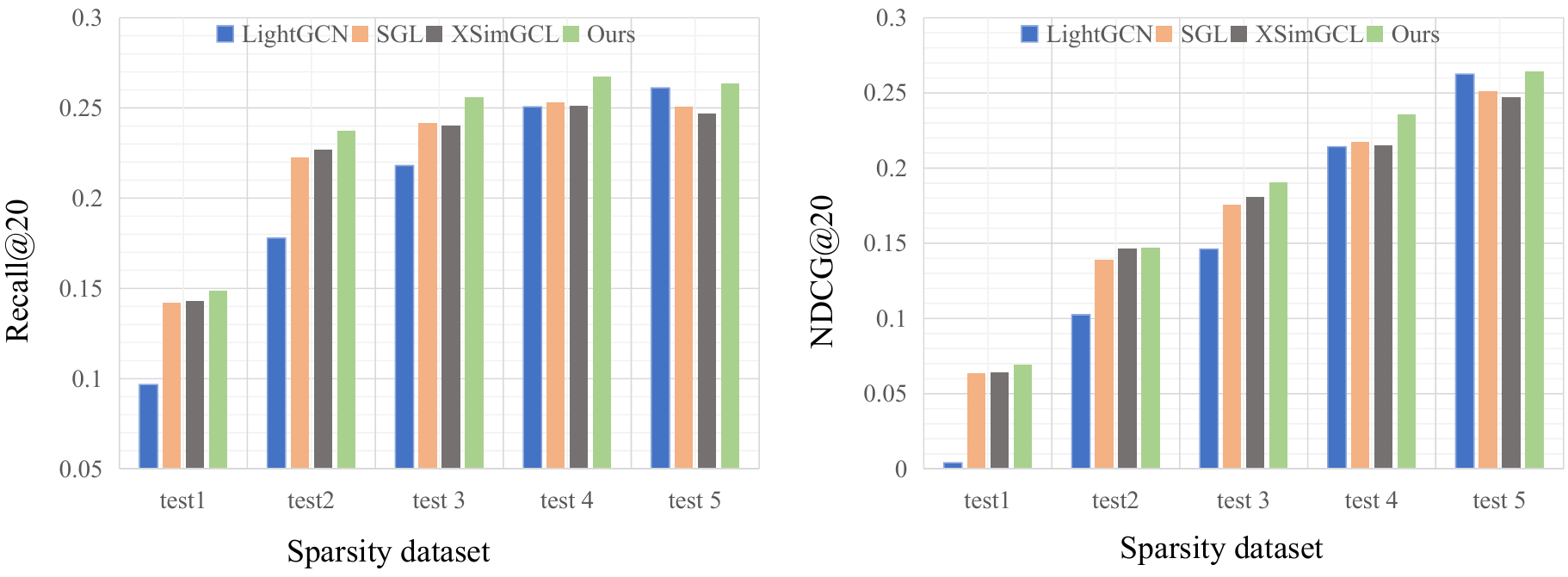}
        \caption{Performance with item interaction numbers}
        \label{fig:sub2}
    \end{subfigure}
    \caption{The performance on data with different sparsity degrees. The sparsity decreases from `test1' to `test5'. The relative improvement of \name in the sparsest case is even larger than the average.} 
    \label{fig:Performance on sparse data}
\end{figure}

\subsubsection{Performance on Sparse Data}.
To verify the effectiveness of \name in handing data sparsity scenarios, we compare it with LightGCN~\cite{he2020lightgcn}, SGL~\cite{wu2021self}, and XSimGCL~\cite{yu2023xsimgcl} in this experiment. Following Yang et al.~\cite{yang2023knowledge},  we divide users and items into 5 groups on average according to the number of interactions in the training data (`test1'->`test5', with decreasing sparsity).
The results on Gowalla are reported in Fig.~\ref{fig:Performance on sparse data}. Similar results are observed on the other datasets and are thus omitted for brevity.

\name achieves the best performance at almost all levels of sparsity. On the `test1' group divided by item with the highest sparsity, \name is 8.5\% higher than SGL and 7.5\% higher than XSimGCL on metric NDCG@20, which is even larger than the average case.
We ascribe this outcome to the presence of our two adversarial perturbators. First, the structure perturbator simulates sparse data scenarios during training, thereby mitigating the detrimental effects of data sparsity on the model. Second, the embedding perturbator is instrumental in constructing more resilient contrastive views, facilitating improved embeddings for users/items. 

\begin{figure}
    \centering
    \includegraphics[width=0.8\textwidth]{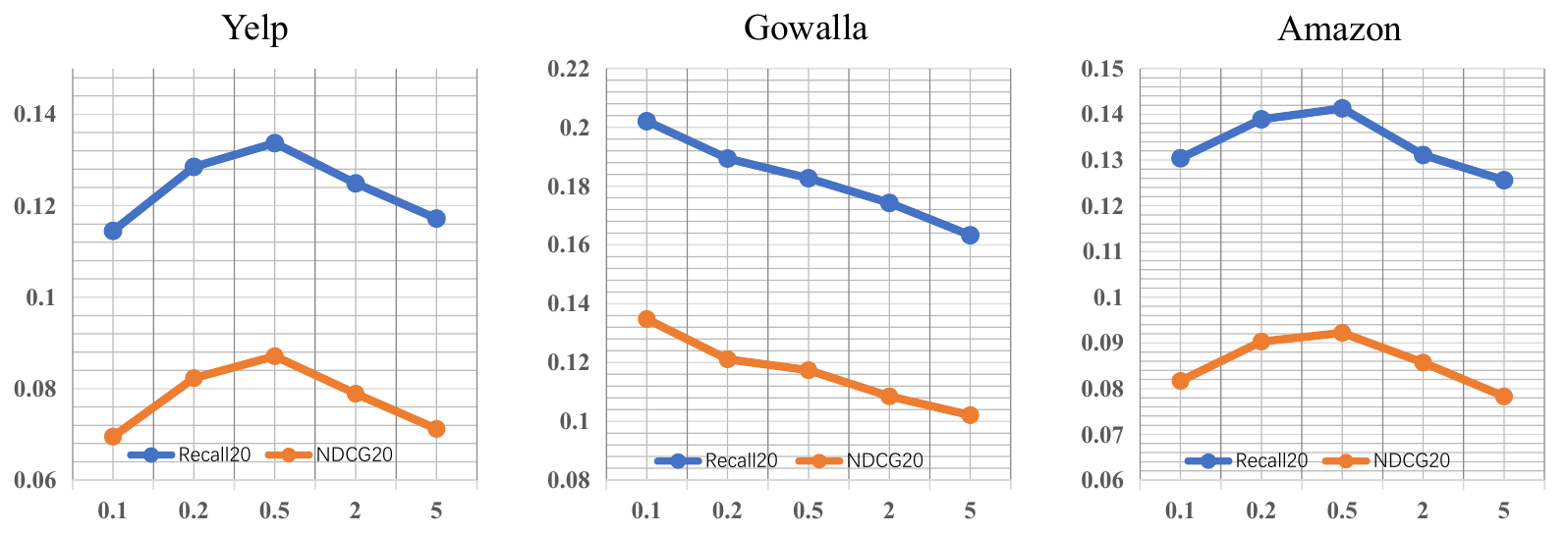}
    \caption{Parameter sensitivity results of $\lambda_1$. The proposed \name performs relatively stable w.r.t. $\lambda_1$ in a wide range.}
    \label{fig:L1 Hyperparameter}
\end{figure}

\subsection{Hyper-parameter Sensitivity (RQ4)}
Finally, we investigate our model’s sensitivity in relation to several key hyper-parameters: the weight $\lambda_1$ for the contrastive loss, the edge selection proportion $\alpha$ in the structure perturbator, and the magnitude $\omega$ in the embedding perturbator. We set the temperature hyper-parameter $\tau$ to be fixed at 0.2, the selected layer for comparison to be the first layer, and the $L_2$ regularization strength $\lambda_2$ to be $1e-4$. The results for $\lambda_1$ are shown in Fig~\ref{fig:L1 Hyperparameter}. The proposed \name performs relatively stable w.r.t. $\lambda_1$ in a wide range. The best result was achieved when $\lambda_1$ was 0.1 on the Gowalla dataset, while Yelp and Amazon achieved the best results when $\lambda_1$ reached 0.5. 

\begin{figure}
    \centering
    \includegraphics[width=0.8\textwidth]{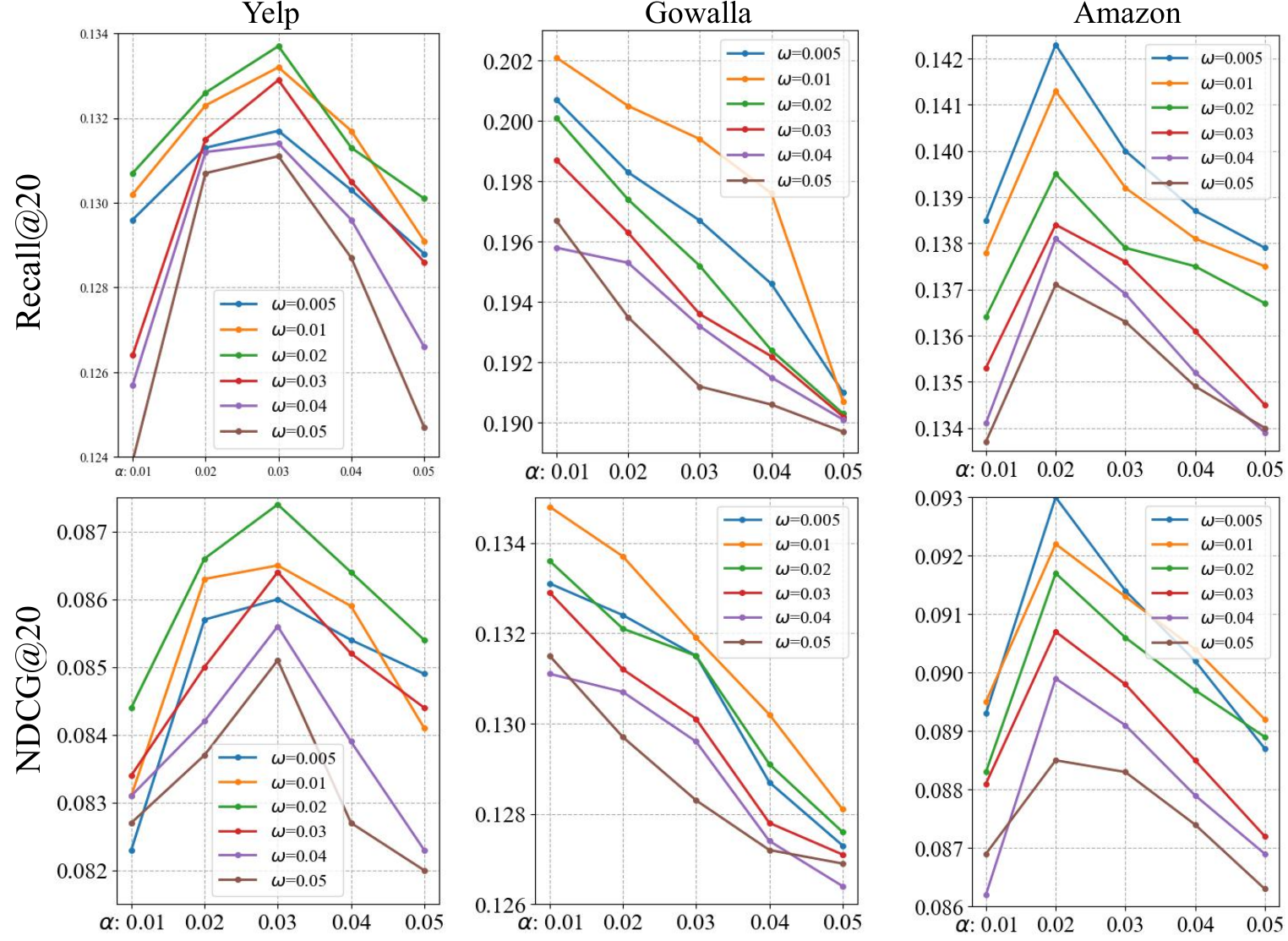}
    \caption{The influence of $\alpha$ and $\omega$. The performance is relatively stable w.r.t. these two hyper-parameters, while the best performance is achieved when both hyper-parameters are set in a modest way.}
    \label{fig:Hyperparameter}
\end{figure}

We next test the two hyper-parameters of $\alpha$ and $\omega$. Here, we take the best performing $\lambda_1$ for each dataset, and try different combinations of $\alpha$ and $\omega$ within the range [0.01, 0.02, 0.03, 0.04, 0.05] and [0.005, 0.01, 0.02, 0.03, 0.04, 0.05], respectively. The results are shown in Fig.~\ref{fig:Hyperparameter}. We found that on the three datasets, the performance is generally better when $\omega$ is relatively small (e.g., $\omega$=0.005, 0.01, 0.02). This suggests that larger perturbation noise can lead to a decrease in model performance. As for $\alpha$, the best results for the three datasets are 0.03, 0.01, and 0.02, respectively. 
In summary, the performance of the model is relatively stable when the three hyper-parameters of \name vary within a wide range.

\section{Related Work}\label{sec:rel}
In this section, we briefly review the related work, including graph contrastive learning for recommendation, graph generative adversarial network.

\noindent{\bf Graph Contrastive Learning for Recommendation}.
In contrast to earlier approaches relying on matrix factorization~\cite{rendle2009bpr,koren2009matrix,liang2018variational,ouyang2014autoencoder}, GCF has become mainstream due to the outstanding performance of GNNs~\cite{wang2019neural,he2020lightgcn,wang2020disentangled,chen2019semi,sun2019multi,chen2020revisiting,yang2021enhanced}. A recent trend of GCF is to incorporate contrastive learning, resulting the so-called GCL-based recommendation~\cite{wu2021self,wei2022contrastive,yu2022graph,lin2022improving,cai2023lightgcl,ye2023towards,yu2023xsimgcl,jiang2023adaptive}. 
Among them, SGL~\cite{wu2021self} introduces structural graph enhancement through the random removal of edges, thereby generating contrasting views. SimGCL~\cite{yu2022graph} opts for embedding space perturbation instead of graph enhancement to establish contrasting views. LightGCL~\cite{cai2023lightgcl} substitutes graph enhancement with the creation of a subgraph via SVD. AdaGCL~\cite{jiang2023adaptive} employs both a graph generative model and a graph denoising model to craft adaptive contrastive views. 

Our work differs from existing studies in two key aspects. First, while existing methods typically construct contrastive views either in the structural space or in the embedding space, \name argues that these two approaches have distinct significance and integrates them into a unified framework. Second, unlike existing methods characterized by high randomness, \name leverages two adversarial perturbators to progressively generate contrastive views with increasing difficulty, a curriculum learning-based approach that facilitates better model performance.

\noindent{\bf Graph Generative Adversarial Network}.
By employing a min-max adversarial framework, GANs~\cite{goodfellow2020generative} have achieved commendable success in a multitude of domains~\cite{madry2019towards,zhang2021inaccurate,metzen2021meta,qian2022integrating}, and graph-related tasks are no exception. For instance, DiGCL~\cite{tong2021directed} proposed the phenomenon that the more difficult it is to contrastive views within a certain range, the better the effect is in the directed graph GCL task. NetGAN~\cite{bojchevski2018netgan} employs random walks to learn a distribution of networks, capturing essential properties of real graphs. GraphGAN~\cite{wang2018graphgan} designs a generator to learn node embeddings and a discriminator to predict link probabilities. ANE~\cite{dai2018adversarial} utilizes GANs as a regularization term to facilitate the learning of robust representations.

More recently, there has been a notable trend in leveraging adversarial learning for the generation of contrastive views, yielding promising results. For example, GASSL~\cite{yang2021graph}  generates challenging contrastive views through the addition of perturbations; ArieL~\cite{feng2024ariel} leverages the idea of information regularization to stabilize training and uses subgraph sampling to achieve scalability; ARIEL~\cite{feng2022adversarial}  incorporates an information regularizer, harnessing Markov relationships between graphs to generate rich contrastive samples; AD-GCL~\cite{suresh2021adversarial} employs a trainable edge deleter and adversarial methods to produce higher-quality contrast views; GACN~\cite{wu2023graph} creates a trainable view generator and jointly trains the GAN model and the GCL model; LDA-GCL~\cite{huang2023adversarial} designs adversarial graph-structured data augmentation to construct contrastive views.  

Compared with the above methods, they almost all generate contrast views by adversarially destroying the original graph structure, while ignoring the significance of vector space. \name uses two lightweight perturbators to achieve double-end adversarial, first extracting low-redundancy subgraphs and then adding adversarial perturbations. This not only avoids the problem of large time and computing resources of traditional GCL methods, but also fills the gap of adversarial perturbations in vector space.

\section{Conclusion}\label{sec:con}
In this paper, we propose a novel GCL-based recommendation model, \name, which generates contrastive views of increasing difficulty through two trainable adversarial perturbators based on the idea of curriculum learning. 
Compared with previous methods that randomly construct contrastive views, our adversarial structure perturbator generates perturbated graphs that are distant from the original user-item interaction graph, and our adversarial embedding perturbator adds parameter perturbations that deviates against the contrastive learning loss. 
Experimental evaluations demonstrate the effectiveness of the proposed approach.
In the future work, we plan to further study the optimization problem over contrastive views and discrete graphs. We are also interested in incorporating extra knowledge into the recommendation process.

\bibliographystyle{unsrt}  
\bibliography{references}

\end{document}